\documentclass[aps,prl,reprint,superscriptaddress,floatfix]{revtex4-1}
\usepackage[english]{babel}
\usepackage{amsmath,amsthm}
\usepackage{amsfonts}
\usepackage{xspace}
\usepackage{bm}
\usepackage[squaren, thinqspace]{SIunits}
\usepackage{booktabs}
\usepackage{graphicx}
\usepackage{color}

\begin{document}
\newcommand{\CSO}{\ensuremath{\mathrm{Cu}_2\mathrm{OSeO}_3}\xspace}
\newcommand{\Hy}{\ensuremath{\bm{H} || \bm{y}}\xspace}
\newcommand{\Nx}{\ensuremath{N_x}\xspace}
\newcommand{\Ny}{\ensuremath{N_y}\xspace}
\newcommand{\Nz}{\ensuremath{N_z}\xspace}


\title{Helimagnon resonances in an intrinsic chiral magnonic crystal}


\author{Mathias Weiler}
\email[]{mathias.weiler@wmi.badw.de}
\affiliation{Walther-Mei{\ss}ner-Institut, Bayerische Akademie der Wissenschaften, Garching, Germany}
\affiliation{Physik-Department, Technische Universit{\"a}t M{\"u}nchen, Garching, Germany}
\author{Aisha Aqeel}
\altaffiliation{present address: University of Regensburg, Regensburg, Germany}
\affiliation{Zernike Institute for Advanced Materials, University of Groningen, Groningen, The Netherlands}
\author{Maxim Mostovoy}
\affiliation{Zernike Institute for Advanced Materials, University of Groningen, Groningen, The Netherlands}
\author{Andrey Leonov}
\affiliation{Zernike Institute for Advanced Materials, University of Groningen, Groningen, The Netherlands}
\affiliation{Center for Chiral Science, Hiroshima University, Japan}
\author{Stephan Gepr\"{a}gs}
\affiliation{Walther-Mei{\ss}ner-Institut, Bayerische Akademie der Wissenschaften, Garching, Germany}
\author{Rudolf Gross}
\author{Hans Huebl}
\affiliation{Walther-Mei{\ss}ner-Institut, Bayerische Akademie der Wissenschaften, Garching, Germany}
\affiliation{Physik-Department, Technische Universit{\"a}t M{\"u}nchen, Garching, Germany}
\affiliation{Nanosystems Initiative Munich, Munich, Germany}
\author{Thomas T. M. Palstra}
\altaffiliation{present address: The University of Twente, Enschede, The Netherlands}
\affiliation{Zernike Institute for Advanced Materials, University of Groningen, Groningen, The Netherlands}
\author{Sebastian T. B. Goennenwein}
\affiliation{Walther-Mei{\ss}ner-Institut, Bayerische Akademie der Wissenschaften, Garching, Germany}
\affiliation{Physik-Department, Technische Universit{\"a}t M{\"u}nchen, Garching, Germany}
\affiliation{Nanosystems Initiative Munich, Munich, Germany}
\affiliation{Institut f\"{u}r Festk\"{o}rper- und Materialphysik, Technische Universit{\"a}t Dresden, Dresden, Germany}
\affiliation{Center for Transport and Devices of Emergent Materials, Technische Universit\"{a}t Dresden, Dresden, Germany}



\date{\today}

\begin{abstract}
We experimentally study magnetic resonances in the helical and conical magnetic phases of the chiral magnetic insulator \CSO at the temperature $T=\unit{5}{\kelvin}$. Using a broadband microwave spectroscopy technique based on vector network analysis, we identify three distinct sets of helimagnon resonances in the frequency range $\unit{2}{\giga\hertz}\leq f \leq \unit{20}{\giga\hertz}$ with low magnetic damping $\alpha\leq0.003$. The extracted resonance frequencies are in accordance with calculations of the helimagnon bandstructure found in an intrinsic chiral magnonic crystal. The periodic modulation of the equilibrium spin direction that leads to the formation of the magnonic crystal is a direct consequence of the chiral magnetic ordering caused by the Dzyaloshinskii-Moriya interaction.  The mode coupling in the magnonic crystal allows excitation of helimagnons with wave vectors that are multiples of the spiral wave vector.
\end{abstract}


\maketitle

Magnons are the fundamental dynamic excitations in ordered spin systems. Spin waves in ferromagnetic materials with collinear magnetic ground state have been a focus of extensive fundamental research~\cite{seavey1958,phillips1966,kalinikos1986}. The field of magnonics deals with the integration of electronics and magnons for data processing applications~\cite{kruglyak2010, lenk2011, demokritov2013,chumak2015}. Key questions and challenges in the field of magnonics relate to dynamics of magnons in laterally confined magnonic waveguides and magnonic crystals~\cite{krawczyk2014}, where magnons can display discrete wavenumbers due to dipolar or exchange interactions~\cite{demidov2009,demidov2015} and the magnon bandstructure can be tailored in analogy to photonic crystals~\cite{john1987,yablonovitch1987}. Magnonic crystals can be artificially created in a \textit{top-down} approach by introducing an \textit{extrinsic} periodic modulation of a magnetic property to an otherwise uniform magnetic crystal or thin film.

Interestingly, materials with chiral magnetic order feature an \textit{intrinsic} modulation of the equilibrium spin direction with periodicity of about \unit{10}{\nano\meter} to \unit{100}{\nano\meter} - most prominently visible in the formation of a skyrmion lattice~\cite{muhlbauer2009}. Hence, such materials should form a natural helimagnonic crystal and provide a \textit{bottom-up} strategy for fabrication of magnonic crystals that go beyond nanolithographic possibilities~\cite{garst2017} by achieving magnetic unit cells in the sub \unit{100}{\nano\meter} range and excellent crystallinity over several millimeters.  Inelastic neutron scattering experiments studied the meV-range bandstructure of \CSO arising due to the crystal lattice constant of about $\unit{0.8}{\nano\meter}$ ~\cite{portnichenko2016, tucker2016}.  Remarkably, the additional magnon bands caused by the finite pitch of about $\unit{60}{\nano\meter}$~\cite{adams2012,seki2012b} are in the GHz frequency range ($<\unit{0.1}{\milli\electronvolt}$), making them inaccessible to inelastic neutron scattering ~\cite{janoschek2010,kugler2015b} but highly relevant for magnonic applications. Magnonic crystals formed by chiral magnets will have great impact on the emerging field of skyrmionics~\cite{schulz2012,nagaosa2013,fert2013}, which aims to exploit individual magnetic skyrmions~\cite{rossler2006,muhlbauer2009,yu2010} for information transport by ultra-low current densities~\cite{jonietz2010,yu2012,jiang2015,woo2016a}.

Here, we provide conclusive experimental evidence for the formation of a magnonic crystal caused by the finite helix pitch formed at low temperatures within a \CSO single crystal by using broadband magnetic resonance spectroscopy. The chiral magnetic insulator \CSO is of particular interest due to its electrically insulating and magnetoelectric properties~\cite{okamura2013a,ruff2015,okamura2015, mochizuki2015a}. Our findings go beyond previous studies of dynamic microwave frequency excitations of chiral magnets~\cite{onose2012,schwarze2015,ehlers2016} and may spark further studies of spin-wave excitation, propagation and quantization in intrinsic chiral magnonic crystals. We furthermore reveal small resonance linewidths of the helimagnons in \CSO that suggest a damping of $\alpha\leq0.003$ at a temperature $T=\unit{5}{\kelvin}$, underpinning the potential merits of \CSO for magnonics and spintronic applications requiring chiral spin-torque materials with low magnetic damping.

\begin{figure}
\includegraphics[width=86mm]{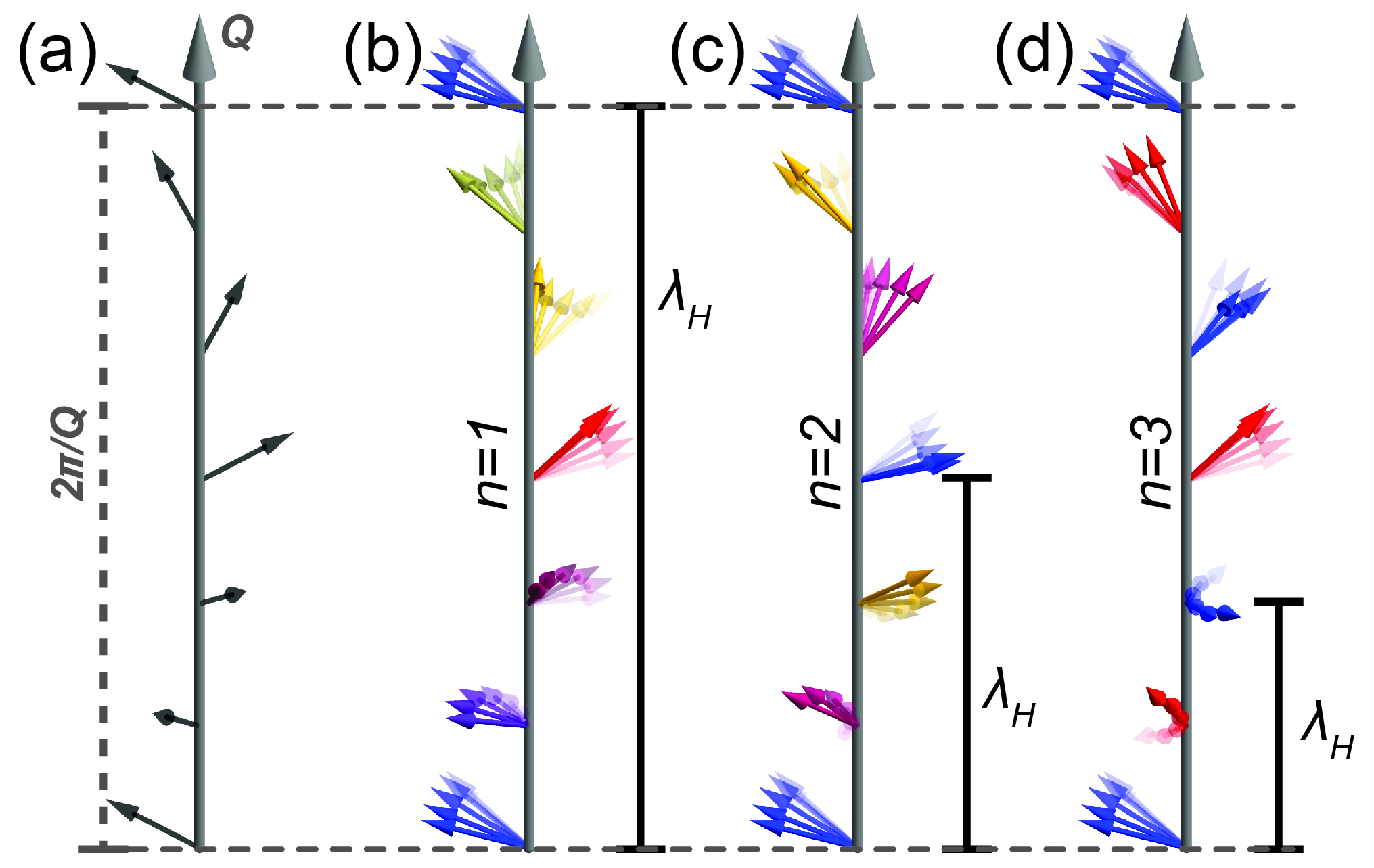}%
\caption{(color online)(a) Equilibrium spin arrangement in the conical phase, showing the helix length $l_\mathrm{H}=2 \pi / Q$. (b) Precessional mode of the $n=1$ conical helimagnon at a snapshot in time. The helimagnon wavelength is $\lambda_\mathrm{H}=l_\mathrm{H}$. The time-evolution of the spin precession is indicated by the transparency of the vectors. The precessional phase depends on the spin position and is represented by the color of the vectors. (c) First higher order ($n=2$) helimagnon with $\lambda_\mathrm{H}=l_\mathrm{H}/2$. (d) Second higher order ($n=3$) helimagnon with $\lambda_\mathrm{H}=l_\mathrm{H}/3$. \label{fig:didactic}}
\end{figure}
The energy of a magnon with a wave vector $\bm{k}$ ($k a \ll 1$, $a$ being  the spin-spin separation) in a ferromagnet with a uniform collinear spin state is  given by $\hbar\omega\approx\hbar\omega_0+D_\mathrm{s} k^2$, where $D_\mathrm{s}$ is the spin wave stiffness and $\hbar$ is the reduced Planck constant. The momentum conservation implies that a spatially  uniform ac magnetic field excites the magnon with $\bm k = 0$ and energy $\hbar \omega_0$ (ferromagnetic resonance).

Dzyaloshinskii-Moriya interaction $D$ transforms the uniform ferrimagnetic state of \CSO into a helical spiral with the wave vector $\bm{Q}$, which in zero field is along one of the cubic axes. Neglecting the small cubic anisotropies, $Q=D/J$~\cite{schwarze2015} with the exchange integral $J$. For \CSO, the pitch is $l_\mathrm{H}=2\pi/Q\approx\unit{60}{\nano\meter}$~\cite{adams2012,seki2012b}. When the applied magnetic field, $\bm{H}_0 \parallel \bm{z}$, exceeds a critical value $H_{\mathrm{c}1}$, the helical spiral turns into a conical spiral with $\bm{Q} \parallel \bm{H}_0$ as shown schematically in Fig.~\ref{fig:didactic}(a). Despite the spatial inhomogeneity of the spiral states, one can still define a conserved magnon wave vector, $\bm{k}$, in the so-called co-rotating spin frame, in which the magnetization vector is constant. The ac magnetic field, which in the co-rotating spin frame has components $\propto e^{\pm \bm{Q} \cdot \bm{z}}$, excites spin waves with $\bm{k} = \pm \bm{Q}$ as depicted in Fig.~\ref{fig:didactic}(b).

The magnetic anisotropy terms allowed by cubic symmetry, such as the quartic magnetic anisotropy $m_x^4 + m_y^4 + m_z^4$, where $\bm{m}$ is a unit vector in the direction of the magnetization, give rise to a non-uniform rotation of spins and add higher harmonics with the wave vectors $n \bm{Q}$ (where $n$ is an integer number) to the spiral. Then the  magnon wave vector is not conserved even in the co-rotating frame. Rather, it becomes a crystal wave vector in the magnonic crystal formed by the distorted spiral where $\bm{Q}$ plays the role of the unit vector of the reciprocal lattice. This leads to formation of magnon bands and opens small gaps in the magnon spectrum. Importantly, since the magnon wave vector is now defined up to a multiple of $\bm{Q}$, the spatially uniform ac magnetic field can excite magnons with the wave vectors $n \bm{Q}$ (wavelength $\lambda_\mathrm{H}=l_{H}/n$). Schematic spin dynamics of the first two higher order modes ($n=2$ and $n=3$) are shown in Figs.~\ref{fig:didactic}(c) and (d), respectively (only the $+nQ$ modes are shown).

Neglecting the changes in the magnon spectrum due to the spiral distortion and the effect of the magnetodipolar interactions which result in the energy splitting of the $n=\pm1$ ($+\bm{Q}$ and $- \bm{Q}$) modes~\footnote{See Supplemental Material [url] for details of data processing, linewidth analysis, skyrmion resonances, magnon spectrum, magnetostatic modes, magnetic anisotropy and electric field excitation, which includes Refs.~[38-39]}\nocite{clogston1956,ruff2015a}, the energy of the magnon with the wave vector $n \bm{Q}$ is~\cite{kataoka1987b, kugler2015b}

\begin{equation}\label{eq:fit}
  \hbar \omega_n=|n| \frac{g \mu_\mathrm{B} B_{\mathrm{c}2}}{1+N \chi} \sqrt{n^2+(1+\chi)\sin^2 \Theta},
\end{equation}
where $\mu_\mathrm{B}$ is the Bohr magneton, $\cos\Theta=\mu_0H_0/B_{\mathrm{c}2}$ is the conical angle, $N$ is the demagnetization factor along the direction of the $\bm{Q}$ vector, $\mu_0$ is the vacuum permeability, $\omega_n=2\pi f_n$ is the angular frequency and
\begin{equation}\label{eq:chi}
  \chi=\mu_0 \frac{M_\mathrm{s}^2}{D Q}
\end{equation}
is the internal conical susceptibility~\cite{schwarze2015} with the saturation magnetization $M_\mathrm{s}$. For $n\neq1$, the energy of the modes of opposite chirality is degenerate also in the presence of dipolar interactions~\cite{Note1}. Note that Eq.~\eqref{eq:fit} does not depend on the sign of $D$, because $Q=D/J$. In the helical phase, the equilibrium orientation of all spins $\bm{S}$ on the helix is $\bm{S}\perp\bm{Q}$ and the net magnetization is zero. This results in a multi-domain state with $\bm{Q}\parallel[100]$ directions~\cite{adams2012}. Under an applied magnetic field, the spiral wave vector may become field-dependent and the evolution of $\bm{Q}$ with $\bm{H}_0$ depends on the domain and the direction of $\bm{H}_0$~\cite{aqeel2016}. In the helical phase, no simple analytical equation for $f_n$ similar to Eq.~\eqref{eq:fit} can be derived. Nevertheless, the helical spiral is a magnonic crystal and the arguments given above concerning the possibility to detect magnon modes with the wave vectors $n \bm{Q}$ still hold. We note that we also expect helimagnon quantization in the skyrmion phase, which can be understood as the superposition of three spin helices at an angle of 120\degree\xspace to each other~\cite{nagao2015}.

To experimentally verify the existence of an intrinsic magnonic crystal resulting in quantized helimagnons in the conical and helical phases of \CSO, we performed broadband helimagnon resonance measurements using a \CSO single crystal cut to a cuboid shape with lateral dimensions $L_x=\unit{4.4}{mm}$, $L_y=\unit{2.0}{mm}$, and $L_z=\unit{0.8}{mm}$. The crystal was grown by a chemical vapor transport method~\cite{belesi2010,aqeel2016a}.
\begin{figure}
\includegraphics{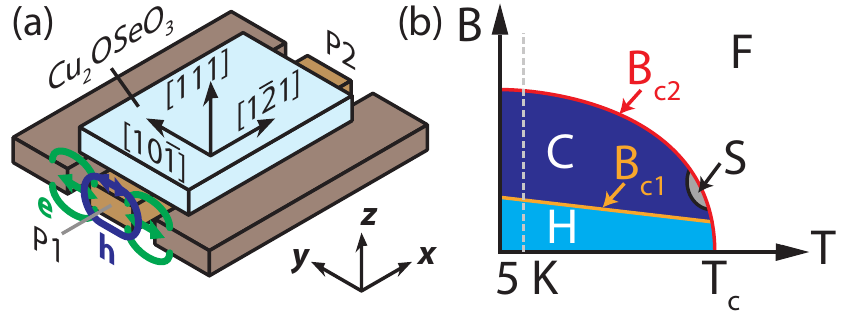}%
\caption{(color online) (a) Sketch of the experimental setup. The (111)-oriented \CSO crystal is placed on top of a CPW. Application of an ac current to the center conductor generates both electric ($\bm{e}$) and magnetic ($\bm{h}$) microwave fields within the sample.  (b) Schematic depiction of the phase diagram of \CSO. H=helical, C=conical, S=skyrmion, F=ferrimagnetic. \label{fig:setup}}
\end{figure}
The \CSO crystal was oriented  using a Laue diffractometer and placed on top of a coplanar waveguide (CPW) with a center conductor width of $\unit{100}{\micro\meter}$ as shown in Figure~\ref{fig:setup}(a). A vector network analyzer (VNA) was connected to the two ports, P1 and P2, of the CPW and the CPW/\CSO assembly was inserted into the variable temperature insert of a superconducting 3D vector magnet. The sample temperature was set to $T=\unit{5}{\kelvin}$ and adjusting the static external magnetic flux density $B=\mu_0H_0$ gave access to the helical (H), conical (C) or ferrimagnetic (F) phases as shown schematically by the dashed line in Fig.~\ref{fig:setup}(b). In all three phases, we excited and detected magnon resonances by measuring the complex transmission S$_{21}$ from P1 to P2 as a function of frequency $f$ with the VNA with fixed microwave power of \unit{1}{\milli\watt} and temperature $T=\unit{5}{\kelvin}$. In our measurements, $\bm{H}$ was applied along $\bm{x}$, $\bm{y}$ and $\bm{z}$ directions and the external magnetic field strength $\unit{-0.3}{\tesla}\leq\mu_0H_0\leq\unit{0.3}{\tesla}$ was swept in increments of $\mu_0\delta H_0=\unit{0.5}{\milli\tesla}$ from positive to negative values.

\begin{figure}
\includegraphics{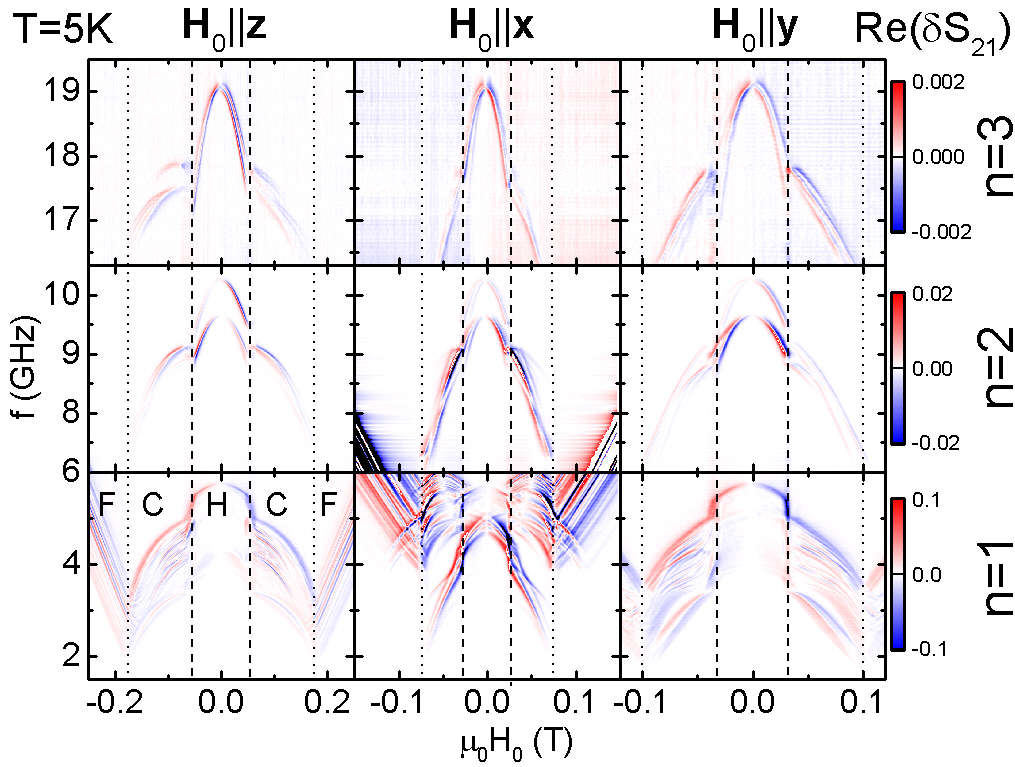}%
\caption{(color online) Colorcoded $\delta S_{21}$ (see text) spectra recorded as a function of $f$ and $H_0$ at T=5K for three different orientations of the external magnetic field $\bm{H}_0$. $H_0$ was swept from positive to negative values. Contrast in $\delta S_{21}$ corresponds to the detection of magnon resonances. Approximate phase transitions are marked by the vertical lines.  \label{fig:colormaps}}
\end{figure}
The normalized field-derivative $\delta S_{21}(f,H_0)$ of $S_{21}$~\cite{Note1, maier-flaig2017d} is shown in Fig.~\ref{fig:colormaps} for all three investigated orientations of $\bm{H}_0$. For clarity, only $\mathrm{Re}\left(\delta S_{21}\right)$ is shown. Contrast in $\mathrm{Re}\left(\delta S_{21}\right)$ is caused by a change of $\partial S_{21}/\partial H_0$ and is attributed to the excitation and detection of spin waves at a frequency $f_\mathrm{res}$. In addition to the previously observed~\cite{onose2012,schwarze2015} resonances at frequencies $f<\unit{6}{GHz}$ (bottom row) in H, C and F phases we also detect helimagnon resonances in both, the C and H phases at higher frequencies (middle and top row). No corresponding resonances are detected in the F phase at these elevated frequencies within the sensitivity of our setup, in agreement with the $\bm{k} = 0$  selection rule in the collinear state. We attribute the three sets of resonances in the C and H phases to the excitation and detection of chiral spin waves with wavelength quantized to integer fractions of the helix pitch $l_\mathrm{H}$ as discussed above. It is remarkable that the $n\neq1$ modes can be excited by our CPW with center conductor width exceeding the pitch by three orders of magnitude. In principle, magnetoelectric coupling allows to excite the $n>1$ modes by the electric field of the CPW, though with vanishingly small efficiency (see~\cite{Note1} for a detailed calculation). Hence, as argued in the context of Fig.~\ref{fig:didactic}, we attribute the excitation of the higher order modes to magnetic anisotropy.

Within the frequency range of our VNA, the $n=4$ mode was not accessible (we anticipate $f_4\approx\unit{30}{\giga\hertz}$). Changing the orientation of $\bm{H}_0$ only quantitatively influences the spectra, the three distinct modes in the H and C phases are always present. We repeated these experiments for $\unit{5}{\kelvin}\leq T \leq \unit{60}{\kelvin}$  and found that the resonances gradually broaden with increasing $T$ such that we could not detect the $n=2$ and $n=3$ modes for $T \gtrsim\unit{20}{\kelvin}$, while the spectra remained qualitatively unchanged. The critical fields $B_{\mathrm{c}1}$ and $B_{\mathrm{c}2}$ of the phase transitions from H to C and C to F phases, respectively, [cf.~Fig.~\ref{fig:setup}(b)] can be deduced from the corresponding discontinuities in $\partial f_\mathrm{res}/\partial H_0$. The thus experimentally determined critical fields are marked by the dashed ($B_{\mathrm{c}1}$) and dotted ($B_{\mathrm{c}2}$) vertical lines in Fig.~\ref{fig:colormaps}.

\begin{figure}
\includegraphics{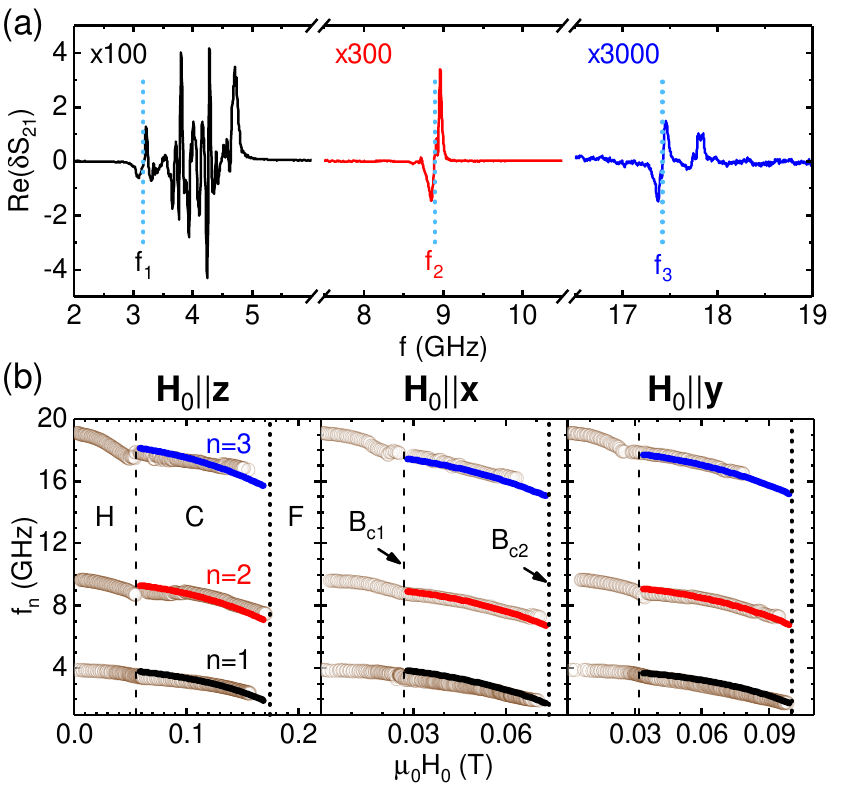}%
\caption{(color online) (a) Raw $\delta S_{21}$ data obtained at $T=\unit{5}{\kelvin}$ with $\bm{H}_0\parallel \bm{z}$ and $\mu_0 H_0=\unit{0.1}{\tesla}$. Vertical dotted lines indicate the resonance frequencies extracted as the lowest-frequency dip-peak zero-crossing of $\delta S_{21}$ for each mode. The data are scaled by the factors above the curves. (b) Experimentally determined resonance frequencies (open circles) plotted as a function of $H_0$ for all three orientations of $\bm{H}_0$. The solid lines represent fits to Eq.~\eqref{eq:fit}. Fitted parameters are given in Table~\ref{tab:fit}.\label{fig:fits}}
\end{figure}
Multiple resonances are observed for all values of $H_0$ for the $n=1$ mode. These multiple resonances continuously connect across the C-F phase transition (see Fig.~\ref{fig:colormaps}) and can be attributed to magnetostatic modes~\cite{walker1957,walker1958,roschmann1977}. We approximately identify the uniform $-Q$ resonance as the mode with the lowest resonance frequency as shown in~\cite{Note1}. Furthermore, the conical $n=1$ modes extend slightly into the ferrimagnetic phase (and vice versa), reminiscent of magnetic soft modes~\cite{montoncello2008,vukadinovic2011}. In both, the helical and conical phases, we observe two sets of resonance for the $n=2$ and $n=3$ modes. This is most easily visible for the $n=2$ helical modes in Fig.~\ref{fig:colormaps} and attributed to a multi-domain state, as previously also observed in the skyrmion lattice phase of \CSO~\cite{zhang2016}. We extract the helimagnon resonance frequencies from Fig.~\ref{fig:colormaps} by determining the zero crossings of $\delta S_{21}(f)$ (corresponding to an abrupt change in the contrast in Fig.~\ref{fig:colormaps}) for each value of $H_0$. A single trace of $\delta S_{21}(f)$ at fixed $\mu_0H_0=\unit{100}{\milli\tesla}$ is exemplarily shown in Fig.~\ref{fig:fits}(a). The vertical dotted lines indicate the extracted resonance frequencies for all $n$.

%
\begin{table}
 \caption{Critical fields, fitted internal conical susceptibility $\chi$ and demagnetization factor $N$ at T=5K for three different orientations of $\bm{H}_0$. The errors represent fit uncertainties. \label{tab:fit}}
  \begin{ruledtabular}
  \begin{tabular}{c c c c}
   & $\bm{H}_0 \parallel \bm{z}$ & $\bm{H}_0 \parallel \bm{x}$ & $\bm{H}_0 \parallel \bm{y}$ \\
  $B_{\mathrm{c}1}$~(T) & 0.055 & 0.027 & 0.032\\
  $B_{\mathrm{c}2}$~(T) & 0.175 & 0.074 & 0.1\\
  $N$ & $0.670\pm0.001$ & $0.089\pm0.001$ &$0.241\pm0.001$\\
  $\chi$ & \multicolumn{3}{c}{$2.76\pm0.01$}\\
  \end{tabular}
 \end{ruledtabular}
\end{table}
The thus extracted resonance frequencies $f_n$ are plotted as open symbols in Fig.~\ref{fig:fits}(b) for $0\leq\mu_0 H_0\lesssim B_{\mathrm{c}2}$. An overlay of these $f_n$ on the data in Fig.~\ref{fig:colormaps} is shown in Fig.~S1~\cite{Note1}. We then performed a global fit of all data in Fig.~\ref{fig:fits}(b) using Eq.~\eqref{eq:fit} for $n=2$ and $n=3$ modes. For the $n=1$ mode we include the effect of sample shape (demagnetization) in our calculations, resulting in a modification of Eq.~\eqref{eq:fit}~\cite{Note1}.  The free fit parameters are $\chi$, $\Nx/\Ny$ and \Nz. We enforce $\Nx+\Ny+\Nz=1$ and use the extracted $B_{\mathrm{c}1}$ and $B_{\mathrm{c}2}$ for each orientation of $\bm{H}_0$. For all fits, we use constant $g=2.1$~\cite{schwarze2015}. The fit is restricted to data obtained in the C phase, where  Eq.~\eqref{eq:fit} is appropriate. The resulting fits are shown by the solid lines in Fig.~\ref{fig:fits}(b). Very good agreement between data and model is achieved and the best fit parameters determined by the Levenberg-Marquardt fitting are summarized in Table~\ref{tab:fit}.  The fitted $\chi=2.76\pm0.01$ is somewhat larger than the previously reported value of $\chi=1.76$~\cite{schwarze2015}. We note that, when fitting data only for a single orientation of $\bm{H}_0$, we actually find $\chi\approx2$ with modified demagnetization factors. Hence, the large value of $\chi$ might be caused by neglecting any further anisotropies (cubic or uniaxial) other than the shape anisotropy. This also explains the slight systematic deviations between the fit and data in Fig.~\ref{fig:fits}(b).  The fitted demagnetization factors $N$ are in excellent agreement with the calculated demagnetization factors for a general ellipsoid of the sample dimensions ($\Nx=0.665$, $\Ny=0.085$, $\Nz=0.250$)~\cite{osborn1945a} and in good agreement with those for a corresponding rectangular prism ($\Nx=0.619$, $\Ny=0.117$, $\Nz=0.265$)~\cite{aharoni1998}. $B_{\mathrm{c}2}$ from Table~\ref{tab:fit} increases for directions with larger demagnetization field due to the decrease of the total conical susceptibility caused by demagnetization~\cite{schwarze2015}.

We experimentally observe helimagnon resonances with low linewidths at $T=\unit{5}{\kelvin}$ in Fig.~\ref{fig:colormaps} and Fig.~\ref{fig:fits}(a). It is hence interesting to extract the damping of the helimagnons in \CSO at this temperature. We carried out a corresponding linewidth analysis of the helical resonances of the $n=2$ and $n=3$ modes for $\bm{H}_0\parallel \bm{z}$ with $\mu_0H_0=\unit{0.1}{\tesla}$ (fits are shown in~\cite{Note1}). Our analysis suggests an upper bound for the  magnetic damping of $\alpha\leq0.003$, which is compatible with the recently reported low-temperature damping in the ferrimagnetic \CSO phase~\cite{stasinopoulos2017}. Due to radiative damping~\cite{schoen2015} or inhomogeneous broadening the actual damping might be even smaller. While still substantially larger than the damping in yttrium iron garnet($\alpha<10^{-4}$~\cite{klingler2017}), the damping in \CSO is comparable to the record value recently reported in a metallic ferromagnetic CoFe alloy at room temperature~\cite{schoen2016}.

We also performed experiments in the skyrmion phase. The data is shown in~\cite{Note1} and allows us to identify clockwise, counterclockwise and breathing modes in agreement with earlier experiments~\cite{onose2012,schwarze2015}. However, we were not able to resolve the higher order modes in the skyrmion phase, presumably due to the much larger linewidths of the magnetic resonances close to $T_\mathrm{c}$.

Taken together, the three distinct sets of resonances observed in Fig.~\ref{fig:colormaps} for each orientation of $\bm{H}_0$ are well described within the simple model given in Eq.~\eqref{eq:fit}. The fits yield parameters for $\chi$ and $N$ that are within the range of expectations. We thus attribute the distinct set of three helimagnon resonances visible for all investigated $\bm{H}_0$ orientations to the experimental observation of the $n=1$, $n=2$ and $n=3$ helimagnon modes of a natural, intrinsic magnonic crystal with low magnetic damping. The naturally formed magnonic crystal in \CSO in conjunction with the low magnetic damping in the helical and conical phases of \CSO opens exciting perspectives for spintronics in chiral magnets. Because chiral magnetic order can be found in many materials with sufficiently large intrinsic or interfacial Dzyaloshinskii-Moriya interaction, including room-temperature systems~\cite{moreau-luchaire2016,takagi2017}, we expect that natural magnonic crystals exist in a wide range of further materials. In addition to temperature, strain~\cite{shibata2015} or doping~\cite{shibata2013} can be used to reconfigure these magnonic crystals.

\begin{acknowledgments}
Financial support from the DFG via SPP 1538 “Spin Caloric Transport” (project GO 944/4 and GR 1132/18) is gratefully acknowledged.
\end{acknowledgments}

\end{document}